\documentclass[prl,aps,twocolumn,showpacs,floatfix]{revtex4}
\usepackage{mathrsfs}
\usepackage{rotating}
\usepackage{color}
\usepackage{graphicx}           
\usepackage{dcolumn}            
\usepackage{bm}                 
\usepackage{hyperref}           
\usepackage[latin1]{inputenc}   
\usepackage{pstricks}           
\usepackage{amsmath,amssymb}
\usepackage{version}
\usepackage{epstopdf}
\DeclareGraphicsExtensions{.eps,.ps}

\usepackage{ulem}
\usepackage{float}
\usepackage{booktabs}
\usepackage{subfigure}
\usepackage{graphicx}
\usepackage{dcolumn}
\usepackage{bm}
\usepackage{ulem}

\begin{document}

\preprint{APS/123-QED}
\title{ Robust Strategies for Affirming Kramers-Henneberger Atoms }
\author{Pei-Lun He}
 \email{a225633@sjtu.edu.cn}
\author{Zhao-Han Zhang}
\author{Feng He}
 \email{fhe@sjtu.edu.cn}
\affiliation{%
Key Laboratory for Laser Plasmas (Ministry of Education) and Department of
 Physics and Astronomy, Collaborative Innovation Center of IFSA
 (CICIFSA), Shanghai Jiao Tong University, Shanghai 200240, China
}%

\date{\today}
\begin{abstract}
Atoms exposed to high-frequency strong laser fields experience the ionization suppression
due to the deformation of Kramers-Henneberger (KH) wave functions, which has not been confirmed yet in any experiment.
We propose a bichromatic pump-probe strategy to affirm the existence of KH states,
which is formed by the pump pulse and ionized by the probe pulse.
In the case of the single-photon ionization triggered by a vacuum ultra-violet probe pulse,
the double-slit character of KH atom is mapped to the photoelectron momentum distribution.
In the case of the tunneling ionization induced by an infrared probe pulse,
streaking in anisotropic Coulomb potential gives rise to the rotation of the photoelectron momentum distribution in the
laser polarization plane.
Apart from bichromatic schemes, the non-Abelian geometric phase provides an alternative route to affirm
the existence of KH states. Following specific loops in laser parameter space, a complete spin flipping transition could be achieved.
Our proposal has advantages of being robust against focal-intensity average as well as ionization depletion,
and is accessible with current laser facilities.
\end{abstract}

\maketitle

Modern light-matter interaction researches date back to Einstein's explanation on the photoelectric effect,
in which ionization happens only if the absorbed photon energy is larger than the binding energy.
The advent of laser technologies has boosted light-matter interaction researches into new eras, where
novel nonperturbative phenomena are discovered, for examples, strong-field tunneling ionization \cite{TI07}, above threshold ionization \cite{ATI},
high-harmonic generation \cite{HHG1,HHG2,HHG3}, nonsequential double ionization \cite{NSDI}, low energy
structures \cite{LES1,LES2}, and photoelectron holography \cite{holo}.
Among these fascinating scenarios, stabilization of atoms in intense laser fields, i.e., the counterintuitive decreasing of
ionization probability with the increasing of driving laser intensities, attracts
attention of the ultrafast community \cite{Gav02,stabs02,Richter}.

Two mechanisms are known for stabilization. One is interference
stabilization \cite{sta1}, in which the released electron wave packets from populated Rydberg states interfere destructively.
The other is adiabatic stabilization, in which the multiphoton ionization
is suppressed due to the deformation of Kramers-Henneberger (KH) wave functions  \cite{Pont,stab1},
which are defined to be the eigenstates of a time-averaged Hamiltonian \cite{KH0}.

Though theoretically predicted for decades, the experimental confirmation of adiabatic stabilization is obscure due to ionization depletion and the focal-intensity average of lasers. In real experiments,
the fine structure related to the stabilization may be smeared out after integrating all ionized fragments
driven by different laser intensities. Furthermore,
while the field strength in the focused center reaches the threshold of
stabilization, the lower intensity around the focusing spot may completely ionize the target.
The target might also be completely ionized before the laser field reaches its peak intensity in the time domain \cite{DV}.
Up to now, there is only tantalizing indirect experimental evidence for the adiabatic stabilization.
For example, in Ref. \cite{evidemce}, a large acceleration of neutral atoms was reported and regarded as a signal of
stabilization \cite{Wei}. However, this evidence is not convincing enough as frustrated ionization \cite{FTI},
in which the ionized electrons get recaptured by the parent nuclei, has a similar output.

There are vast researches on adiabatic stabilization \cite{Gav02,stabs02,Richter}.
However, only a few attempted to directly identify KH states.
Kulander \textit{et al.} suggested that the appearance of the even order of high-harmonic generation \cite{KHHHG} is
a manifest of KH states. Morales \textit{et al.} identified specific fine structures in photoelectron momentum distribution contributed by
excited KH states  \cite{Smirnova}.
Jiang \textit{et al.} suggested that the photoelectron momentum distribution carrying dynamical interference structures
provides information on adiabatic stabilization \cite{Jiang}.
However, these proposals are sensitive either to the laser intensity or to the pulse envelope and are
not robust against ionization depletion. Thus, the experimental realization is still challenging.

In this letter, we proposed to detect KH states using a bichromatic pump-probe strategy, in which the KH state is formed by the pump
pulse and ionized by the probe one. By detecting the photoelectron momentum distribution, one is able to extract the dichotomic structure
of the target, and thereby affirm the existence of KH states. The spin flipping for atoms following a loop in the laser parameter space
provides an alternative route.

\begin{figure}
\centering
\includegraphics[width=0.5\textwidth]{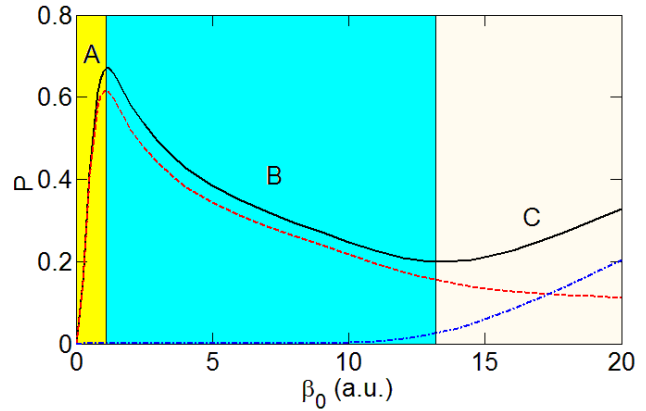}
\caption{ \label{fig1}
The ionization probability as a function of $\beta_0$ obtained from TDSE simulations.
The black solid line is for total ionization, the red dashed line is for one-photon ionization probability,
and the blue dotted-dash line describes the nonadiabatic ionization.
}
\end{figure}

Our start point is the three-dimensional time-dependent Schr\"odinger equation (TDSE) in the KH frame \cite{KH0} (atomic units are used throughout
unless stated otherwise)
\begin{equation}
i \frac{\partial}{\partial t} \psi(\mathbf{r}, t)=  \left[ \frac{1}{2}\mathbf{p}^{2}+V(\mathbf{r}+\boldsymbol{\beta}) \right ] \psi(\mathbf{r}, t),
\label{sch}
\end{equation}
where $\boldsymbol{\beta}$ is the time-dependent electron displacement
$\boldsymbol{\beta}=\boldsymbol{\beta}_0 \sin(\omega_0 t)$, with $\boldsymbol{\beta}_0=\beta_0 f(t)\mathbf{e}_x$ and $\beta_0=\frac{E_0}{\omega^2}$
if the driving laser field is linearly polarized.
The corresponding laser field is given by
$\mathbf{E}(t)=-\frac{\partial^2}{\partial t^2}\boldsymbol{\beta}$.
We use the envelope $f(t)=\cos^2(\pi t/L) \ (-L/2<t<L/2)$ throughout this paper, where $L$ stands for the pulse duration.
The ground state of Eq. (\ref{sch}) is obtained using the imaginary time method \cite{imag},
and the split-operator method is adopted to propagate the wave function in real time.
By Fourier transforming the ionized wave function,
we obtained the photoelectron momentum distribution.
Using the hydrogen atom as the prototypical target, we calculated the ionization probability as a function of
$\beta_0$, as shown by the black solid line in Fig. \ref{fig1} (a). Here, the laser pulse has a frequency of $\omega_0=3 $ a.u.,
and a total duration $L$ of sixty cycles.
The ``death valley" \cite{DV} structure is clearly shown.

Researches on different aspects of high-frequency-laser ionization scatter in references \cite{dich,omg2,KHT2,DI1,DI2,Rost17,Jiang2,Toyota,Rost18,Wei18}, for our purpose here, we
summarize the main conclusions with a special emphasis on the role played by KH states.
We expanded the oscillating Coulomb
potential as $V(\textbf{x}+\boldsymbol{\beta})\approx\sum_\text{n} V_\text{n}(\textbf{x};\boldsymbol{\beta}_0)e^{-i \text{n} \omega_0 t}$ \cite{enve,Rost15}.
$V_0(\textbf{x};\boldsymbol{\beta}_0)$ provides a laser-dressed adiabatic potential, while the nonzero harmonic components
$V_\text{n}(\textbf{x};\boldsymbol{\beta}_0)$ induce photon absorption/emission. The Hamiltonian in Eq. (\ref{sch}) can now be regrouped into two
parts, i.e., the adiabatic term $H_0=\frac{\textbf{p}^2}{2}+V_0(\textbf{r};\boldsymbol{\beta}_0)$ and the remaining part
$H_I=\sum_{\text{n} }V_\text{n}(\textbf{x};\boldsymbol{\beta}_0)e^{-i \text{n} \omega_0 t}$.
The time-dependent electron wave packet can be expanded as
\begin{equation}
\Psi(t)=\sum_{N}C_N(t)e^{-i\int^{t}d\tau E_N(\tau)}\phi_N(\boldsymbol{\beta}_0),
\label{Psi}
\end{equation}
in which $\phi_N(\boldsymbol{\beta}_0)$ is the instantaneous eigenstate of $H_0$ and satisfies the governing equation
$H_0\left |\phi_N(\boldsymbol{\beta}_0)  \right \rangle=E_N(\boldsymbol{\beta}_0)\left |\phi_N(\boldsymbol{\beta}_0)  \right \rangle$.
Inserting Eq. (\ref{Psi}) into the Schr\"odinger equation yields
\begin{eqnarray}
\begin{matrix}
\dot{C}_N=\sum_M \left \langle N \right| \left (-\frac{\partial \boldsymbol{\beta}_0 }{\partial t}  \frac{\partial }{\partial \boldsymbol{\beta}_0} -iH_I\right )\left  |M \right \rangle \\
 \times e^{-i\int^{t}d\tau (E_M-E_N)} C_M.
\end{matrix}
\label{Cn}
\end{eqnarray}
The term $ \left \langle N \right|H_I\left |M \right \rangle$ is responsible for photons absorption/emission,
and the skew-hermitian matrix $\left \langle N \right| \frac{\partial }{\partial \boldsymbol{\beta}_0} \left |M \right \rangle $ provides
the nonadiabatic coupling \cite{KHT2,Toyota,Rost18} and the geometric phase \cite{geo0}.
As indicated by Eq. (\ref{Cn}), KH states are of central importance here. The deformation of the KH wave function $\phi_N(\boldsymbol{\beta}_0)$
leads to the suppression of $\left \langle N \right|V_\text{n}\left |M \right \rangle$, which is the fundamental reason for adiabatic
stabilization \cite{Pont,stab1}. The phase accumulation due to the distorted KH state \cite{Wei} leads to the dynamic
interference \cite{DI1,DI2,Rost17,Jiang2}. Furthermore, KH states determine the strength of nonadiabatic coupling
$\left \langle N \right| \frac{\partial }{\partial \boldsymbol{\beta}_0} \left |M \right \rangle $.

With Eq. (\ref{Cn}) in hand, we explored scenarios of ionization shown in Fig. \ref{fig1} by dividing
the ionization probability curve into three stages marked by A, B, and C. In stage A, $\beta_0$ is small and so is
the changing rate $\frac{\partial \boldsymbol{\beta}_0} {\partial t}$, which
means the deformation of KH wave functions and the nonadiabatic coupling are negligible.
We extracted the single-photon ionization
fragment from the total ionization spectra and presented it by the red dashed curve in Fig. \ref{fig1}.
The single-photon-ionization probability overlaps with the
total ionization probability as $\beta_0<1$ a.u., which suggests
that the ionization can be well described by the conventional first-order perturbation theory and the nonadiabatic coupling is negligible.
In stage B, the total ionization probability decreases due to the significant deformation of KH wave functions.
The norm of $\left \langle N \right|V_\text{n}\left |M \right \rangle$ is suppressed with an increasing $\beta_0$.
The dichotomic characteristic of KH states, i.e., the dimensionless number $\frac{Z}{\beta_0 I_p}$,
serves as a measure of the deformation of wave functions.
For the ground state hydrogen atom, the nuclear
charge $Z=1$ a.u. and the ionization energy $I_p=0.5$ a.u.. $\frac{Z}{\beta_0 I_p }\approx 1$ roughly corresponds
to the point where the second order derivative of the laser-dressed ground state energy curve vanishes \cite{supp}.
$\frac{Z}{\beta_0 I_p}<1$ implies the deformation of the wave function is significant.
In stage C, the nonadiabatic ionization becomes more and more important \cite{KHT2,Toyota,Rost18}. There is no distinct
boundary between B and C. The nonadiabatic coupling is determined by the product of
$\left\langle N\left|\frac{\partial}{\partial \boldsymbol{\beta}_0}\right| M\right\rangle$ and $\frac{\partial \boldsymbol{\beta}_0}{\partial t}$, which implies an envelope dependence.
Similar to the excited-state tunneling ionization \cite{STI}, in the nonadiabatic ionization, the atom first transits
to excited states, which are then mediated to continuum states. The effective ionization potential decreases
with the increasing of $\beta_0$, as shown in the supplementary file \cite{supp}, and thus the
nonadiabatic ionization is more important for a larger $\beta_0$.
Ionization from excited states dominates since the ionization potential gets smaller in this situation.

With these understandings about the central role played by KH states in the $\omega_0 \gg I_p$ regime, we now make the following proposal to experimentally
affirm KH states.
The strategy is basically a pump-probe scheme: a
linearly or circularly polarized high-frequency strong laser pulse is used to irradiate on a prototypical hydrogen atom,
which is to be ionized by another circularly polarized probe pulse. When the electron is released from the KH hydrogen atom,
who plays the role of a double-slits, the photoelectron momentum distribution will inherit
the double-slit interference structure. In principle, this proposal can already be performed on advanced laser facilities \cite{LF1,LF2,LF3}.
In this strategy, the probe pulse should be strong enough to trigger noticeable ionization,
but not so strong that the formed KH state gets destroyed. This imposes a constraint
$I_2 \ll (\frac{\omega_2}{\omega_1})^4I_1 $, where $I_1$ and $I_2$ are intensities of the pump and probe pulses.
Note that the subscript 0 is preserved for the case of using only one pulse. Besides that, laser frequencies of the pump and probe pulses
should be proper so that the photoelectron momentum distributions induced by the pump and probe pulses do not overlap.
$\omega_2$ should be sufficiently large to avoid interfering with very low energy electron produced by the nonadiabatic coupling \cite{KHT2,Toyota}.


\begin{figure}
\centering
\includegraphics[width=0.5\textwidth]{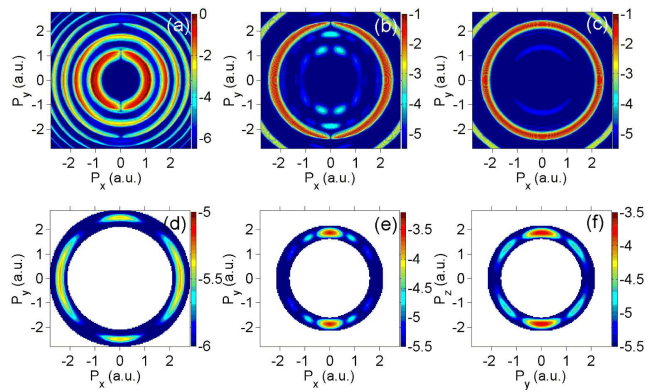}
\caption{ \label{fig2}
Upper row: The photoelectron momentum distributions contributed by both the pump and probe pulses.
Lower row: The photoelectron momentum distributions contributed by the one-probe-photon ionization,
which are picked out from the upper row. Different laser parameters are
used for the three columns. Left column: $\beta_1=2$ a.u., $\omega_1=1$ a.u., and $\omega_2=3.5$ a.u.;
The pump pulse is linearly polarized along the $x$ axis with a duration of twenty optical cycles,
and the probe pulse is circularly polarized in the $x-y$ plane.
Middle column: $\beta_1=5$ a.u., $\omega_1=3$ a.u., and $\omega_2=2$ a.u.;
The pump pulse is linearly polarized along the $x$ axis with a duration of sixty optical cycles,
and the probe pulse is circularly polarized in the $x-y$ plane.
Right Column: $\beta_1=3$ a.u., $\omega_1=3$ a.u., and $\omega_2=2$ a.u.;
The pump pulse is circularly polarized in the $x-y$ plane with a duration of sixty optical cycles,
and the probe pulse is circularly polarized in the $y-z$ plane.
In all panels, the probe pulse has a duration of ten optical cycles and intensity $I_2= 10^{16}$ $\text{W}/\text{cm}^2$.
The delay between the pump and probe pulses is zero.
}
\end{figure}

The upper row of Fig. \ref{fig2} shows the above-threshold-ionization (ATI) containing the fragments released by absorbing
$n \omega_1$ photons and $m \omega_2$ photons, where $n$ and $m$ are integers.
Though the probability of absorbing $\omega_2$ is small due to the relatively weak intensity of the probe pulse,
the ionization by the probe pulse contributes distinguished angular distribution and non-overlapping photoelectron energy
with the ionization fragments induced by the pump pulse. Thus one
can easily separate one-probe-photon ionization from the dominating pump-photon ionization,
as shown in the lower row in Fig. \ref{fig2}. The laser parameters for the three
columns are presented in the caption. The ionization amplitude of the formed KH states, with the field parameters used
in (d), is proportional to $J_n(p_x A_1/\omega_1)J_{m+1}(\sqrt{p_x^2+p_y^2} A_2/\omega_2)+J_n(p_x A_1/\omega_1)J_{m-1}(\sqrt{p_x^2+p_y^2} A_2/\omega_2)$, where $J_n$ is the $n$-th order
Bessel function \cite{supp}. All panels in the lower row present clearly angular nodal structures, which are the manifestation
of double-slit interference \cite{DoubleS}. The number of nodes is determined by $\beta_1\sqrt{2E_\textbf{k}}$ with $E_\textbf{k}$ the photoelectron energy.
Inversely, $\beta_1$ can be extracted from the
angular distribution of the photoelectron induced by the one-probe-photon ionization. The comparison of (a), (b) and (c) shows that a
larger $\omega_1$ is more convenient for separating the ionization events from the pump and probe pulses. A larger frequency $\omega_1$ is also better for avoiding the ionization depletion \cite{omg}.

\begin{figure}
\centering
\includegraphics[width=0.5\textwidth]{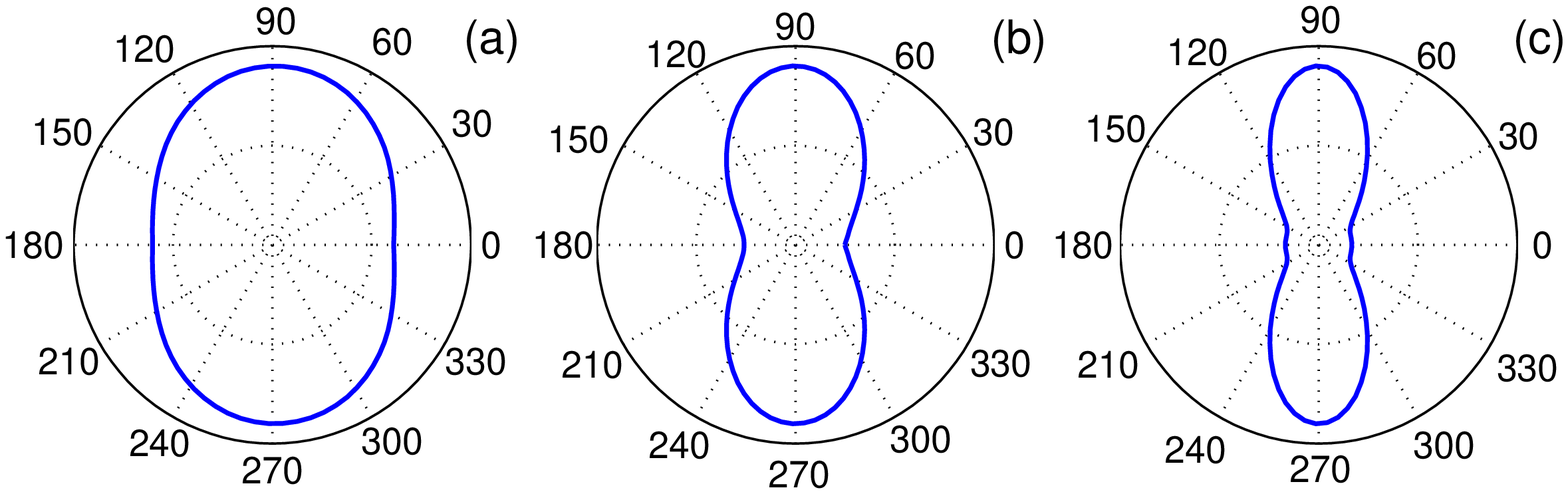}
\caption{ \label{fig3}  Focal-intensity averaged photoelectron angular distributions.
The laser frequencies  are $\omega_1=1$ a.u., and $\omega_2=3.5$ a.u.. The pump laser intensity is
(a) $I_1=3.5\times10^{16}$ W/cm$^2$ ( $\beta_1=1$ a.u. ), (b) $I_1=1.4\times10^{17}$ W/cm$^2$ ( $\beta_1=2$ a.u. ),
and (c) $I_1=3.2\times10^{17}$ W/cm$^2$ ( $\beta_1=3$ a.u.).
The probe laser has the intensity $I_2=1\times10^{16}$ W/cm$^2$, and the duration of ten optical cycles.
}
\end{figure}

The unavoidable focal-intensity average in real experiments must be taken into account to judge the feasibility of the above proposal.
By assuming that the intensity of the laser pulses has the spatially Gaussian distribution \cite{intavg},
we plotted the focal-intensity-averaged photoelectron angular distribution in Fig. \ref{fig3}.
The frequencies are $\omega_1=1$ a.u. and $\omega_2=3.5$ a.u.,
corresponding to the parameters used in Fig. \ref{fig2} (a)(d).
The laser intensities used for the three columns from left to right are $I_1=3.5\times10^{16}$ W/cm$^2$ ( $\beta_1=1$ a.u. ),
$I_1=1.4\times10^{17}$ W/cm$^2$ ( $\beta_1=2$ a.u. ), and $I_1=3.2\times10^{17}$ W/cm$^2$ ( $\beta_1=3$ a.u. ), respectively.
The photoelectron from unperturbed hydrogen atoms should be rotational invariant in the laser polarization plane, hence, the
anisotropic ionization probability, as shown in all these panels, confirms the existence of dichotomic distribution of the KH hydrogen atom, which in turn provides the evidence
of adiabatic stabilization.
The onset of adiabatic ionization implies that ionization is mainly contributed from small $\beta_1$ \cite{supp},
which explains why interference structures in photoelectron momentum distributions are not as distinct as that in Fig. \ref{fig2}.
As the number of nodes is determined by $\beta_{1} \sqrt{2 E_{\textbf{k}}}$, a larger $\omega_2$ is favored to produce distinctive angular distributions.
This strategy works for diverse laser parameters and is robust against focal-intensity average.
Moreover, the pump-probe delay can be tuned thus the probe pulse can contribute noticeable ionization before the KH atom is depleted.

\begin{figure}
\includegraphics[width=0.5\textwidth]{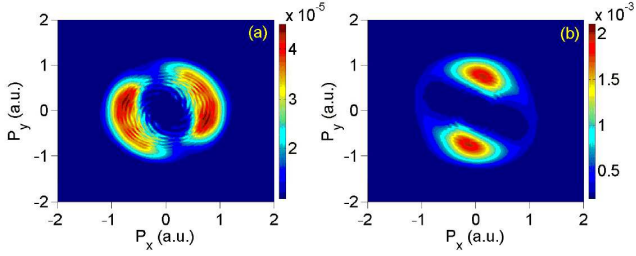}
\caption{ \label{fig4} Focal-intensity-averaged photoelectron momentum
distributions that are tunneling ionized by infrared probe pulses.
The pump and probe laser parameters are (a) $\beta_1=2$ a.u. and $I_2=3\times10^{14}$ W/cm$^2$,
and (b) $\beta_1=4$ a.u. and $I_2=2\times10^{14}$ W/cm$^2$.
The probe pulse has {\blue a} duration of four optical cycles.
}
\end{figure}

Instead of the single-photon ionization triggered by the high-frequency probe pulse, the KH atom may be tunneling ionized
by an infrared probe pulse. Figure \ref{fig4} shows the focal-volume-averaged photoelectron momentum
distributions for the probe laser intensity (a) $I_2=3\times10^{14}$  W/cm$^2$ and $\beta_1=2$ a.u. and
(b) $I_2=2\times10^{14}$  W/cm$^2$ and $\beta_1=4$ a.u.. Here, He$^+$ in the KH state is prepared and used as the target, and the two cases have the same Keldysh parameter.
In contrast to the one-probe-photon ionization, signals from large $\beta_1$ are dominating in the tunneling regime \cite{supp}.

The streaking of the photoelectron momentum distribution in the anisotropic Coulomb field produces a tilt angle, which is a function of $\beta_1$ and $I_2$ \cite{tilt1,tilt2}.
The existence of $\beta_1$ can thus be mapped into the streaking tilt angle. We point out that it is also possible to use the laser induced electron diffraction \cite{LIED1,LIED2} or charge resonance enhanced ionization \cite{CRI} to reconstruct $\beta_1$ \cite{supp}.

Besides utilizing the double-slit interference structure and the anisotropic angular streaking, the electron spin-flipping provides another route to affirm KH atoms.
The physical principle is based on the non-Abelian geometric phase. Neglecting the ionization of KH atoms
for a moment, the nonadiabatic coupling among those non-degenerate states, i.e.,
\begin{eqnarray}
\left \langle N \right| \frac{\partial }{\partial \boldsymbol{\beta}_0} \left |M \right \rangle =\left \langle N \right| \frac{\partial H_0}{\partial \boldsymbol{\beta}_0} \left |M \right \rangle/(E_M-E_N),
\label{degen}
\end{eqnarray}
can be suppressed in the adiabatic limit.
However, situations are different for systems with energy degeneracies, where non-Abelian geometric
phases play a role \cite{geo1}.
Denoting the polarization axis of the laser pulse by $\textbf{n}=(\sin\theta\cos\phi,\sin\theta\sin\phi,\cos\theta)$,
we have degenerate KH states $\left|\pm\left(\theta,\phi \right)\right\rangle$ satisfying $\textbf{n}\cdot\mathbf{J}
\left|\pm \left(\theta,\phi \right)\right\rangle=\pm1/2\left|\pm \left(\theta,\phi \right)\right\rangle$,
where $\mathbf{J}$ is the addition of orbital angular momentum $\mathbf{L}$ and spin angular momentum $\mathbf{S}$.
$\left|\pm \left(\theta,\phi \right)\right\rangle$ are connected to $\left|\pm\left(\theta=0,\phi=0\right)\right\rangle$ via rotations
\begin{equation}
 \left|\pm \left(\theta,\phi \right)\right\rangle=\exp(-iJ_z \phi)\exp(-iJ_y \theta)\left|\pm\left(\theta=0,\phi=0\right)\right\rangle.
\end{equation}
Components of non-Abelian connection 1-form $\mathfrak{A}=i\left\langle M |\textbf{d}| N \right \rangle$, where $\textbf{d}$ is exterior derivative, are thus given by
\begin{eqnarray}
\begin{matrix}
\mathfrak{A}_\theta=\frac {1} {2} \begin{pmatrix}
0 & -i \kappa\\
i \kappa & 0
\end{pmatrix}, ~
\mathfrak{A}_\phi=\frac 1 2\begin{pmatrix}
\cos\theta & -\sin\theta \kappa\\
-\sin\theta \kappa & -\cos\theta
\end{pmatrix}.
\end{matrix}
\end{eqnarray}
Here $\kappa=2 \left\langle +\left(\theta=0,\phi=0\right) \right|J_x\left| - \left(\theta=0,\phi=0\right)\right \rangle $ is a real valued function of $\beta_0$, up to a trivial phase factor. We have $\kappa=1$ when $\beta_0\rightarrow 0$.
In the limit $\beta_0\rightarrow \infty$, $\kappa= 0$ if we choose $\left|+\left(\theta=0,\phi=0\right)\right\rangle=\left|L_z=1,S_z=-1/2\right\rangle$ and $\left|-\left(\theta=0,\phi=0\right)\right\rangle=\left|L_z=-1,S_z=1/2\right\rangle$. Such a choice is possible here, as the off-diagonal elements of spin-orbital coupling matrix vanish.

Consider the situation that the laser field is linearly polarized along the $z$ axis. One
adiabatically rotates the polarization axis by $\theta_0=\pi/2$ in the $z-x$ plane, then rotates it by $\phi_0$
in the $x-y$ plane, and finally rotates it back to the $z$ axis.
The holonomy $W=P\left \{ e^{i\oint{\mathfrak{A}}} \right \}$ for such a closed path is
\begin{widetext}
\begin{equation}
W=\begin{pmatrix}
\cos \left(\frac{\kappa \phi_0}{2}\right)+i \sin \left(\frac{\kappa \pi}{2}\right) \sin \left(\frac{\kappa \phi_0}{2}\right) & -i \cos \left(\frac{\kappa \pi}{2}\right) \sin \left(\frac{\kappa \phi_0}{2}\right)\\
-i \cos \left(\frac{\kappa \pi}{2}\right) \sin \left(\frac{\kappa \phi_0}{2}\right) & \cos \left(\frac{\kappa \phi_0}{2}\right)-i \sin \left(\frac{\kappa \pi}{2}\right) \sin \left(\frac{\kappa \phi_0}{2}\right)
\end{pmatrix}.
\label{W}
\end{equation}
\end{widetext}
By setting $\phi_0=\pi/\kappa$, we have the spin flipping transition amplitude $-i \cos \left(\frac{\kappa \pi}{2}\right)$, which is zero when $\kappa=1$ and maximal when $\kappa$ approaches zero.

The deformation of KH states is crucial here.
When the laser field is not very strong, i.e. $\beta_0\ll1$ a.u., the dynamical evolution of the TDSE is determined by the
dipole coupling matrix $\left \langle N \right| \textbf{r} \left |M \right \rangle$. Typically, only a minor fraction of spin flipping could be achieved \cite{nondipole}.
However, with an increasing $\beta_0$, couplings with the highly oscillating laser pulses are suppressed, which reduces Eq. (\ref{Cn}) into
\begin{eqnarray}
\begin{matrix}
\frac{\partial C_N }{\partial  \boldsymbol{\beta}_0}\approx-\sum_{E_M=E_N} \left \langle N \right| \frac{\partial }{\partial  \boldsymbol{\beta}_0} \left |M \right \rangle  C_M
\end{matrix}
\label{Cna}
\end{eqnarray}
in the adiabatic limit.

The spin flipping in KH atom due to the adiabatically rotating electric field is isomorphic to that of in diatomic molecule due to the rotating molecule axis \cite{geo2}.
The nontrivial value $\kappa \neq 1$ manifests the breaking of atomic isotropic symmetry thus the existence axial symmetric KH atom.
In this strategy, ensuring the adiabaticity implies that $\frac{\partial \boldsymbol{\beta}_0 }{\partial t}$  should be small.
Therefore, a pulse with a very long duration is demanded. In this case, in order to avoid ionization depletion,
a laser pulse with {\blue a} large $\omega_0$ is required.

To summarize, a pump-probe scheme is suggested to detect the KH state. The pump pulse creates a KH atom,
whose dichotomic structure is imprinted on the photoelectron momentum distribution. This strategy is
robust against the focal-intensity average and ionization depletion. Alternatively, the spin flipping
induced by the non-Abelian geometric phase in the adiabatically changing laser field can also provide the evidence of KH atoms.
An ideal implementation of our proposal requires $\frac{\omega_1}{I_p}>1$ and $\frac{Z}{\beta_1 I_p}<1$.
In our strategy, the required laser intensity is about
$I_1= 10^{16}$ -- $10^{17}$ $\text{W}/\text{cm}^2$, and the laser wavelength is around 10 -- 50 nm,
which is within the reach of current laser facilities.
Relativistic and quantum electrodynamics effects \cite{QED1,QED2} are negligible for the considered laser parameters.
Moreover, we are concerned only about one-probe-photon ionization from KH states, thus the deviation from relativistic theory mainly differs by scaling factors \cite{Dirac}.
We observed that the low energy electron ionized by the nonadiabatic coupling is altered by the nondipole effect \cite{Smeenk,Ludwig}.
However, this does no harm to our proposal due to the non-overlapping energy.
Our schemes not only provide accessible routes for detecting KH states, thus adiabatic stabilization,
but are also useful for understanding up-coming high frequency strong laser-matter interaction.

This work was supported by National Key R\&D Program of China (2018YFA0404802),
Innovation Program of Shanghai Municipal Education Commission(2017-01-07-00-02-E00034),
National Natural Science Foundation of China (NSFC) (Grant No. 11574205, 11721091, 91850203), and Shanghai Shuguang Project (17SG10).
Simulations were performed on the $\pi$ supercomputer at Shanghai Jiao Tong University.

\end{document}